\documentclass[10pt]{article}
\usepackage{epsfig}
\newcommand{\beao}{\begin{eqnarray*}}
\newcommand{\eeao}{\end{eqnarray*}}
\newcommand{\be}{\begin{equation}}
\newcommand{\ee}{\end{equation}}
\newcommand{\bea}{\begin{eqnarray}}
\newcommand{\eea}{\end{eqnarray}}
\newcommand{\beq}{\begin{eqnarray}}
\newcommand{\eeq}{\end{eqnarray}}
\newcommand{\nn}{\nonumber}

\newcommand{\la}{\lambda}

\newcommand{\Ref}[1]{(\ref{#1})}

\begin{document}
\title{Spontaneous symmetry breaking \\ in the O(4) scalar model on a lattice}
\author{
 V.~Demchik \thanks{Email: vadimdi@yahoo.com},
 A.~Gulov \thanks{Email: alexey.gulov@gmail.com},
 V.~Skalozub \thanks{Email: skalozubv@daad-alumni.de}\\~\\~
 ~{\small \sl Dnipropetrovsk National University, Dnipropetrovsk, Ukraine}
}
\maketitle

\begin{abstract}
The spontaneous symmetry breaking at zero temperature in the four-component four-dimensional scalar $\lambda \phi^4_4$ model (the O(4) model) is investigated on a lattice for different values of the coupling constant $\lambda$. A general method for dealing with such type dependence is developed. The Goldstone modes are integrated out in the spherical coordinates in the internal space of the scalar field by the saddle point method, and the initial functional integral of the model is reduced to an effective one-component theory convenient for lattice investigations. The partition function of this theory is calculated analytically in the static limit and demonstrates the $\lambda$-dependence which is characteristic for distinguishing symmetry breaking. Monte Carlo simulations are performed with the QCDGPU software package on a HGPU cluster. It is shown that the symmetry is spontaneously broken for $\lambda>\lambda_0 \simeq 10^{-5}$. For smaller coupling values, the scalar field vanishes on a lattice, which can be interpreted as instability of the homogeneous condensate or even instability of the model itself. The critical value $\lambda_0$ is independent of the lattice sizes $16^4$ and $32^4$ investigated.
\end{abstract}

\section{Introduction}
Symmetry breaking and phase transitions in the $O(N)$ scalar models are the problems of great importance in  quantum field theory. They are investigated in numerous publications for different number of components $N$ and different space-time dimensions $d = 2,3,4$ by using various methods of calculations, in particular,  Monte Carlo (MC) simulations on a lattice  (see \cite{ZinnJustin:1996cy}-\cite{Marko:2013lxa} and references therein). In what follows, we fix the space-time dimensionality $d = 4$ and consider the  models at zero temperature (symmetric lattices). The numbers of components $N = 1$ and $N > 1$ correspond to qualitatively different physics related with spontaneous breaking of discrete and continuous symmetry. The choice $N = 4$ has relevance to either the One-Higgs-Doublet standard model of particles or low energy limits of QCD and its phase structure at finite temperature. The number $N = 8$ corresponds to the Two-Higgs-Doublet standard model which is a possible candidate for substituting the standard one. Other values of $N$ and $d$ also find physical applications.

Recently, it was discovered in MC simulations for the $O(1)$, $d = 4$ model that the type of the phase  transition at finite temperature depends on the value of coupling constant $\lambda$ \cite{Bordag:2012nh}. This  phenomenon had not been discussed in the literature before. Usually, it is assumed that the coupling is small and has the order $\la \sim 0.01 -  0.1$, whereas the type of the phase transition changes at much lower values of the coupling. Naturally, this problem is of interest for other $O(N)$ models. However, for $N\geq 2$ computations become much more complicated in comparison with the one-component model. In general, the question whether or not spontaneous breaking of continuous symmetry takes place for different $d = 1, 2, 3, 4$ is discussed for many years. We can also mention other related problems as the `triviality', dependence on the number of spatial dimensions, the Goldstone theorem on a lattice, etc. In fact, one needs a reliable order parameter distinguishing the phases and dependence on the coupling values. Traditional investigations of this problem is based on  the effective potential approach (see \cite{ZinnJustin:1996cy}-\cite{Cea02}).

In the present paper, we make a  step in this direction and investigate spontaneous symmetry breaking (SSB) in the $O(4)$ model on a lattice at zero temperature and its dependence on the value of  $\lambda$. The first main problem is a correct treatment of the Goldstone modes. These modes are related with spontaneous breaking of continuous internal symmetries. So, it is impossible to realize them on a lattice exactly, only some remnants can be observed. The second problem is to work out a procedure for MC simulations which can be efficient for an extremely wide interval of coupling values.  To clarify the first problem, we consider the behavior of the partition function (PF) in the continuous theory in case when the homogeneous condensate of scalar field forms the background for quantum fluctuations. We develop a general approach for solving this problem by using the representation of the PF in the spherical coordinates in the internal space of scalar fields. The main idea is to integrate out the continuous angular modes, which become Goldstone bosons after symmetry breaking, before lattice investigations. As a result, we relate SSB with the presence of the scalar field condensate. In such a way we derive an effective action for the radial field $R(x)$.  With this effective action, either symmetry breaking or its dependence on $\lambda$ value are investigated in the way similar to the $O(1)$ case \cite{Bordag:2012nh}. To solve the second problem, we introduce special dimensionless variables in sect. 6.

The paper is organized as follows. In the next section, we define the $O(4)$ model in the spherical coordinates. In sects.~3 and 4 we develop the integration procedure for this case and introduce the PF. Then, in sect.~5 we calculate analytically the PT and investigate its $\lambda$-dependence. In sect.~6 we carry out MC simulation for the radial field obtained after integration over the angular modes and investigate how the SSB depends on the value of coupling $\lambda$. The last section is devoted to discussions of the results obtained and prospects for future studies.

\section{The model}
In this section we describe the $O(4)$ $\phi^4$-model in terms of the spherical variables.  The Lagrangian reads
\be
\label{Lagrangian} L =  \frac{1}{2} \frac{\partial \phi}{\partial
x_\mu} \frac{\partial \phi}{\partial x_\mu} - V(\phi), \ee where
\be \label{V} V(\phi) =   -\frac{m^2}{2}\phi^2  + \la \frac{\phi^4}{4} \ee
and $\phi^2 = \sum_i \phi_i \phi_i$, $i = 1,...,4$. For $m^2 \ge 0$ the SSB potentially takes place.

Let us parameterize the scalar field in the spherical coordinates as $\phi_i = R(x) n_i(x)$, where $n_i$ is a direction in the internal space, $\sum_i n_i n_i = 1$. The direction $n_i$ contains angular variables describing a point on the four-dimensional sphere,
\bea \label{angulars}
n_1 = \sin \theta_1 \sin \theta_2 \sin \theta_3, &&n_2 = \sin \theta_1 \sin \theta_2 \cos \theta_3, \\ \nn
n_3 = \sin \theta_1  \cos \theta_2, &&n_4 =  \cos \theta_1.
\eea
In these variables, the Lagrangian \Ref{Lagrangian} has the form
\be \label{Lagrangian1} L =  \frac{1}{2} \frac{\partial R}{\partial x_\mu} \frac{\partial R}{\partial x_\mu} - \frac{1}{2} R^2 n_i\frac{\partial^2 n_i}{\partial x_\mu \partial x_\mu}  - V(R) \ee
up to the total derivative $\partial {(R^2 n_i \partial n_i/\partial x_\mu )/ \partial x_\mu}$. We also take into account the condition
\be \label{dn2}
\frac{\partial ( n_i n_i)}{\partial
x_\mu} = 0.
\ee
From  \Ref{Lagrangian1} the equations of motion follow:
\bea \label{eqni} &&R^2(x) \frac{\partial^2 n_i}{\partial x_\mu \partial x_\mu} + 2 R(x) \sum_i n_i  \frac{\partial n_i}{\partial x_\mu}  \frac{\partial R}{\partial x_\mu}= 0, \\ \nn
 \label{eqR} &&\frac{\partial^2 R(x)}{\partial x_\mu \partial x_\mu} + \frac{ \partial V(R)}{\partial R} + 2 R \sum_i\frac{\partial n_i}{\partial x_\mu} \frac{\partial n_i}{\partial x_\mu}= 0. \eea
The last terms can be transformed further using Eq.~(\ref{dn2}). As a result, we obtain the equations of motion
\bea \label{eqniF}
&&R^2(x) \frac{\partial^2 n_i}{\partial x_\mu \partial x_\mu} = 0, \\ \nn
 \label{eqRF}
&&\frac{\partial^2 R(x)}{\partial x_\mu \partial x_\mu} + \frac{ \partial V(R)}{\partial R} -2 R \sum_i n^i \frac{\partial^2 n_i}{\partial x_\mu \partial x_\mu}= 0. \eea
At $R = \mathrm{const}$, $n_i$ describes massless particles corresponding to the Goldstone modes appearing after symmetry breaking.

Our next step is to calculate the PF in the spherical coordinates in some way. As we mentioned in  Introduction, our goal is to take into consideration the contributions of the continuous Goldstone modes analytically. After that, the radial field $R(x)$ remains the only dynamical variable. In this way we  obtain an effective  theory convenient for further lattice investigations.  In what follows, we will use the saddle-point approach for integration. It well recommends itself, in particular, in describing the SSB. This is because its  leading approximation accounts for the non-analytic correlations existing between the parameters of the problem. Remaining contributions can be calculated  in perturbation theory. Another advantage of this method is the possibility to extend the integration contour from a saddle point position to infinity reducing the complicated integration in the spherical coordinates to usual Gaussian integrals.  In the next section, we compute an ordinary Gaussian integral in these coordinates. This will demonstrate how the functional integral of interest can be computed.
\section{Saddle-point integration in the spherical coordinates}
When the Cartesian coordinates are used in the internal space, the functional integrals in the $O(N)$ model are of the Gaussian type for free particles. In the spherical coordinates, the Jacobian appears as an additional factor in the integrand. This requires  developing  a procedure for calculating such integrals. As a first step, we compute ordinary Gaussian integrals in spaces with different dimensions.

First, we consider the two-dimensional integral
\begin{eqnarray}
I_2=\int_{-\infty}^{\infty}dx\, dy\ e^{-ax^2} e^{-ay^2}=\frac{\pi}{a}
\end{eqnarray}
and calculate it in a way convenient for further generalization to functional integrals. We write
\begin{eqnarray}
I_2=\int_0^\infty R\,dR\int_0^{2\pi}d\theta\ e^{-aR^2}.
\end{eqnarray}
To calculate this integral, we apply the saddle-point method. Namely, we integrate over the angular variable $\theta$ and transform the integrand into the form
\begin{eqnarray}\label{Demchik:I2}
I_2=2\pi\int_0^\infty dR\ e^{-aR^2+\log R}.
\end{eqnarray}
The stationary points are determined as extrema of the function $F(R)=-aR^2+\log R$ standing in the exponential,
\begin{eqnarray}
\frac{\partial F(R)}{\partial R}=-2aR+\frac{1}{R}=0.
\end{eqnarray}
Hence, we find the saddle point $R_0 = 1/\sqrt{2a}$, and the function can be expanded at this point: $F(R) = -1/2 - 2a(R - R_0)^2 + ...$.  Then, Eq.~(\ref{Demchik:I2}) takes the form
\begin{eqnarray}
I_2=2\pi\frac{e^{-1/2}}{\sqrt{2a}}\int_0^\infty dR e^{-2a(R-R_0)^2}.
\end{eqnarray}
In the saddle-point technique, we stretch the integration contour near the saddle point to the infinite line and
obtain the Poisson integral,
\begin{eqnarray}
I_2=2\pi\frac{e^{-1/2}}{\sqrt{2a}}\sqrt{\frac{\pi}{2a}}=\frac{\pi}{a}\sqrt{\frac{\pi}{e}}.
\end{eqnarray}
This result is very close to the exact value $\pi/a$. The difference can be decreased if the next-to-leading terms of the asymptotic expansion are taken into consideration.

This simple example demonstrates how to proceed in case of integrals in the spherical coordinates. The angular integration is carried out explicitly. After that, the integration is reduced to standard saddle-point calculations with the function $F(R)$ including the exponentiation of the Jacobian. Due to the Jacobian, the stationary point in the radial direction is never located at the origin $R=0$, where singularities may appear.

Let us also adduce the value of the integral obtained in the leading approximation for the four-dimensional case:
\begin{equation}
I_4=\left(\frac{\pi}{a}\right)^{2}\cdot \left(\frac{3^{3/2}\sqrt{\pi}}{e^{3/2}}\right).
\end{equation}
Again, this result is close to the exact value given by the first factor.

From these examples it becomes clear how to proceed in case of functional integrals in the spherical coordinates and calculate the PF.

\section{The partition function}

The PF of the $O(4)$ model in terms of spherical coordinates in the internal space reads
\begin{eqnarray} \label{Demchik:Z}
Z &=& \int D \phi\ \exp\left(\int d^4 x L \right)\nonumber \\
&=& \int_0^{\infty}R^{{3}} D R(x)\int  d^{4 - 1} \Omega(x)  \exp \left( \int d^4 x L \right),
\end{eqnarray}
where $d^{4 - 1} \Omega(x)$ represents the integration over angular variables. Useful information on the free field case as well as a general theory on continual integration is given in numerous books (see, for example, \cite{Shvartz:1993}). In this section, we work out a procedure for calculation in the form convenient for lattice investigations and related to Ref. \cite{Bordag:2012nh}. That is, we assume that a lattice in the Euclidian space-time is introduced with the spacing $a$. The field $R(x)$ is determined by its values in each site $x$ of the lattice.  In what follows, we consider zero temperature and use symmetric lattices.

We consider the constant solution $R_c = \mathrm{const}$, $\theta_i = \mathrm{const}$ as the background for quantum fluctuations and calculate the corresponding PF. A usual procedure in this case is to calculate the effective potential $V(\phi_c)$ and apply the saddle-point method. For constant fields, the equations to find the stationary point are simplified significantly, and the integration in the functional integral can be easily fulfilled.

The PF can be written in the form
\begin{eqnarray} \label{Demchik:Z1}
Z = \prod_x\int_0^\infty D R(x) \int d^{4-1}\Omega(x)\ e^{3\log R(x)}e^{a^4{L}(x)}.
\end{eqnarray}
The `effective potential' for this case is
\begin{eqnarray}\label{Demchik:EPVR}
\tilde{V}(R)=\sum_x a^4 \left(\frac12 m^2 R^2-\frac{\lambda}{4}R^4\right)+\sum_x (3 \log R).
\end{eqnarray}
In fact, this is not a usual effective potential obtained by the Legendre transformation but a useful tool for calculation by the saddle-point method.

The stationary equation, $\partial \tilde{V}/\partial R = 0$, reads
\begin{eqnarray}
R^4-\frac{m^2}{\lambda}R^2-\frac{3}{\lambda a^4}=0.
\end{eqnarray}
It has one real solution
\begin{eqnarray}\label{Demchik:y0}
R_0^2=\frac{m^2}{2\lambda}\left(1+\sqrt{1+\frac{12\lambda}{m^4 a^4}}\right).
\end{eqnarray}
For small $\lambda$, it can be written as $R_0^2 =m^2\lambda^{-1}+3m^{-2} a^{-4}$. The second derivative in the vicinity of $R_0^2$ equals to
$\tilde{V}^{\prime\prime}(R_0)=-\kappa^2=-\left(12\lambda m^{-2}+3m^{-2} a^{-4}\right)$. Hence, the effective potential can be presented as
\begin{eqnarray}
\tilde{V}(R)=\tilde{V}(R_0)-\frac12 \kappa^2\rho^2,
\end{eqnarray}
where
\begin{eqnarray}
\tilde{V}(R_0)=\log \left(\frac{m^2}{\lambda}+\frac{3}{m^2 a^4}\right)^{3/2}+\frac{m^4 a^4}{4\lambda}-
\frac{9}{4}\frac{\lambda}{m^4 a^4}
\end{eqnarray}
and $\rho = R(x)- R_0$. After that, the integration over the field $\rho$ can be fulfilled even with accounting for the kinetic term in the action. But it does not matter at this moment because we observe the important nonanalytic behavior of the functional integral over $\rho$ -- it is proportional to
\begin{eqnarray}\label{Demchik:z0}
Z_0= \left(\frac{m^2}{\lambda}+\frac{3}{m^2 a^4}\right)^{3/2}\exp\left(\frac{m^4 a^4}{4\lambda}\right)
\exp\left(-\frac{9}{4}\frac{\lambda}{m^4 a^4}\right),
\end{eqnarray}
and the two first factors are diverging at $\lambda\to 0$. This is  distinguishable feature of the SSB.  It is worth to mention that such a singular behavior is
similar to that for the mean scalar field calculated from the classical field equations. However, now it comes at quantum level of calculations. On the other hand, in the symmetric phase (for the `normal' mass term, $m^2<0$ in our notations), the integral is proportional to $(3m^{-2} a^{-4})^{3/2}$ and exhibits regular
$\lambda$-dependence in the limit $\lambda\to 0$.

The factor $Z_0$ stands independently of the operator structure of the kinetic part. So, this property of $Z$ can be used to determine the symmetry breaking as a function of $\lambda$. Due to the described behavior, the PF can serve as an order parameter for symmetry breaking in the $O(N)$ models. This behavior could  change  when  other terms standing in the functional integral are taken into consideration. However, we expect that the SSB is expressed in the divergent behavior of the PF in the limit $\lambda\to 0$. Other details on the features  of $Z$ will be discussed in what follows. Of course, the PF itself is not an observable object in actual (for instance, perturbative) calculations. But nevertheless its properties influence the behavior of other observables.

We also have calculated the one-loop corrections to \Ref{Demchik:z0} and convinced  that the first factor is canceled in the total but the second one, which is singular in the limit $\lambda \to 0$, remains unchanged.
\section{Angular integrations and the effective one-component model}
Now, we carry out angular integration in the internal space of the $O(4)$ model. The functions $n_i$ are given in \Ref{angulars}. For the second term in \Ref{Lagrangian1} we get
\be \label{L2O3} L(\theta_i) = \frac{1}{2} R^2(x)[(\partial_\mu \theta_1)^2 + (\partial_\mu \theta_2)^2 \sin^2 \theta_1 + (\partial_\mu \theta_3)^2 \sin^2 \theta_1 \sin^2 \theta_2], \ee
and the Jacobian reads
\be \label{JO4} J(\theta_i) = R^3(x) \sin^2 \theta_1 \sin \theta_2. \ee

To calculate the PF, we construct the `effective potential' \Ref{Demchik:EPVR} and determine the stationary point $R_0$ for the fixed direction $\theta_1 = \theta_1^{(0)}, \theta_2 = \theta_2^{(0)}, \theta_3 = \theta_3^{(0)}$.  Correspondingly, we write the integration variables as $R(x) = R_0 + \rho(x),~
\theta_i(x) = \theta_i^{(0)}+ \eta_i(x)$, $i = 1,2,3.$ After that, we can integrate over angular variables $\eta_i(x)$ by using the saddle-point method.

Substituting the angles in the Jacobian $\sin \theta_i(x) = \sin (\theta_i^{(0)}+ \eta_i(x)), i = 1, 2$, we expand it in series over $\eta_i(x)$: $\sin (\theta_i^{(0)}+ \eta_i(x))= \sin\theta_i^{(0)} + \cos \theta_i^{(0)} \eta_i + O( \eta_i^2) .$ In the given approximation, the angular term \Ref{L2O3} becomes
\be \label{exponential}
L(\theta_i) = - \frac{R_0^2 a^4}{2} [ \eta_1 \partial^2_\mu \eta_1 + \eta_2 \partial^2_\mu \eta_2 ~\sin^2 \theta_1^{(0)}+~ \eta_3 \partial^2_\mu \eta_3 ~\sin^2 \theta_1^{(0)} \sin^2 \theta_2^{(0)}].\ee
Then, we can extend the integration limits for $\eta_i$ to plus-minus infinities and make the change of   variables: $r_1 = (R_0 a^2) \eta_1, ~r_2 = (R_0 a^2) \eta_2 \sin\theta_1^{(0)},$  $r_3 = (R_0 a^2) \eta_3 \sin\theta_1^{(0)} \sin \theta_2^{(0)}$. After these transformations, we obtain the PF measure \Ref{Demchik:Z}:
\be \label{Jac1} R^3(x) \sin^2 \theta_1(x)\sin \theta_2(x)\ D R\, D \theta_1 D \theta_2D \theta_3 = \frac{R^3(x)}{R_0^3a^6}\ D \rho\, D r_1 D r_2 D r_3, \ee
and the angular exponential contains $- \frac{1}{2} \sum_{i} r_i \partial^2_\mu r_i  $. Due to symmetry of the integration limits, the linear in $\eta_i$ terms do not contribute to the result. Thus, we obtain three integrals over the angular variables. These integrals are independent of any physical variables and, after extending the limits of integration to infinities, result in some constant factor. This factor does not influence the PF behavior in the limit $\la \to 0$. Another  property of Eq. \Ref{Jac1} is independence of the r.h.s. of  the arbitrary saddle point parameters $\theta_i^{(0)}$.  The dependence is canceled, as it is expected. Hence, we conclude that the SSB of continuous symmetry can be completely related with the radial variable $R(x)$. After angular integrations, we can integrate over the radial field $\rho$.

It is also worth to note that angular modes are massless in both the restored and broken phases. In the former case all modes are massless whereas after symmetry breaking the massless Goldstone particles have to exist. Hence, it is reasonable to integrate over all the angular modes in the same manner  and derive the effective theory applicable in both the phases. Other  observation is that angular integrations result in a general numeric factor which is independent of $\la$ and therefore unimportant.  This general feature makes the accuracy of these integrations not very essential  for the problem under consideration.

In lattice calculations, we can start just from the obtained result. The factor $R^3(x)/R_0^3$ in Eq.\Ref{Jac1} can be transformed to the extra term in the `effective action' \Ref{Demchik:EPVR} for the one-component field $R(x)$. This observation gives a possibility for formulating a general procedure for investigations of the phase transitions in the $O(N)$ models. Namely, we can start with the effective one-component Lagrangian consisting of the initial
one written in the spherical coordinates where we can omit the angular terms and add the term $\log (R/R_0)^{(N - 1)}$. The value of the saddle point $R_0$ has to be calculated from the `effective potential' $\tilde{V}(R)$ for fixed angular variables.  The procedure of dealing with the one component field $R(x)$ can be simply realized on  lattices similarly to the $O(1)$ case \cite{Bordag:2012nh}.

In what follows, we restrict ourselves to investigation of the SSB at zero temperature, only.  To realize that we take two symmetric lattices with the sizes $16^{4} $ and $32^{4}$ and suppose a homogeneous field condensate in some fixed direction in the internal space as strong background for quantum fluctuations. Different lattice sizes allow to control finite-volume effects. Then  we try to see the phase with the broken symmetry. If we obtain results from MC simulations, which disagree with the hypothesis of quantum fluctuations over the condensate background, we conclude that there is no broken phase at zero temperature. Although we have no direct order parameter after the integration over the Goldstone modes in our approach, we can still measure the mean value of the radial field (the mean absolute value of vector $\phi_i$ in the initial internal space). This quantity cannot vanish in a self-consistent picture of the broken phase. Thus, our strategy is a proof by contradiction, and the statement to check is: the broken phase is realized at zero temperature.

To complete this part we stress once again that the integration over angular modes in the spherical coordinates gives a possibility for a consistent treatment of symmetry breaking on the basis of the effective theory for the radial variable $R(x)$. In this approach,  the role  of the continuous modes related with  compact coordinate lines results in a general numeric  factor which is independent of $\la$  and therefore inessential.

\section{Monte Carlo lattice simulations}
The phase with the broken symmetry in the $O(N)$ model at zero temperature on a lattice can be described by the Euclidian effective one-component Lagrangian consisting out of the radial part of the initial action and the additional term $\log (R/R_0)^{(N - 1)}$ from the Jacobian and integrated angular part:
\begin{equation}\label{Demchik:lataction0}
S =  \sum _x a^4\left(
\frac{1}{2} \frac{\partial R}{\partial x_\mu} \frac{\partial R}{\partial x_\mu} - \frac{m^2}{2}R^2  + \la \frac{R^4}{4}
\right)
- \sum _x \log (R/R_0)^{(N - 1)},
\end{equation}
where $a$ is the lattice spacing, and the value $R_0$ is calculated from the `effective potential' $\tilde{V}(R)$ for fixed angular variables.
Let us remind that integration over continuum compact variables is carried out explicitly in the saddle-point technique as it is described in the previous sections. For MC simulations, we use a symmetric hypercubic lattice with hypertorous geometry in the four-dimensional space-time.
The one-component non-negative scalar field $R(x)$ is defined in lattice sites. As it is known, symmetric $d=4$ lattice corresponds to zero temperature, whereas finite temperatures require a less number of sites in the temporal direction than in the spatial directions.

Since we are interested in varying $\lambda$ in a wide interval of values, self-adjust\-ment of the lattice spacing and other parameters of the model is of great importance. In this regard, we rewrite the action through dimensionless quantities. The classical condensate $R_c=m/\sqrt{\lambda}$ can be set as a unit of the scalar field. There is also the dimensionless parameter
\begin{eqnarray}
z=\frac{\lambda}{m^4 a^4}.
\end{eqnarray}
Writing the radial field as
\begin{eqnarray}
\frac{R(x)}{R_c}=\sqrt{2\Phi_x},
\end{eqnarray}
we obtain the lattice action (up to a constant)
\begin{eqnarray}\label{Demchik:lataction}
S &=& \frac1z\sum _x \Phi_x\left(\Phi_x-1+\frac14\sqrt{\frac{z}{\lambda}}
\sum_\mu\log^2\frac{\Phi_{x+a_\mu\hat{\mu}}}{\Phi_x}\right)
\nonumber \\&&
-\sum_x
\frac{N-1}{2}\log \Phi_x,
\end{eqnarray}
where the log-squared term encodes the kinetic part of the action written through the finite differences instead of derivatives.

The effective model (\ref{Demchik:lataction0}) is derived by integration of angular degrees around the condensate direction. This means quantum fluctuations have to be not larger than  the values of  the condensate.  So, the values $R\gg R_c$ could be incompatible with the effective model. Fortunately, the probability of such large values decreases as $\exp(-\Phi^2/z)\simeq \exp(-R^4/(4zR_c^4))$, $R\to\infty$, in accordance with (\ref{Demchik:lataction}). In this regard, the interval of values of $R$ appears to be actually finite and can be cut by some upper bound $R_\mathrm{max}$. In MC simulations, we use two values $R_\mathrm{max}=3.75R_c$ and $R_\mathrm{max}=4R_c$ to show that the results are independent of the cutoff scale. In the selected interval, $R$ is taken to be uniformly distributed in accordance to the definition of the partition function of the effective theory. The chosen cuts correspond to probability $< 10^{-22}$ (the choice of $z$ is described below), so the obtained Boltzmann ensembles demonstrate no essential dependencies on $R_\mathrm{max}$.

Parameter $z$ is related to the saddle point position $R_0$ defined in (\ref{Demchik:y0}), $(R_0/R_c)^2=(1+\sqrt{1+4(N-1)z})/2$. We use $z\sim 1$ ($z=0.67$ and $z=1$, actually) corresponding to saddle point position at the central part of the interval of field values, $1<R_0/R_c\le 1.5$.
Extremely small values of $z$ could make the $N$-dependent term in the effective action negligible with respect to other terms, whereas large $z$ could move the saddle point outside the cutoff scale.

\begin{figure}
\centering
\includegraphics[bb=1 218 633 623, width=.8\textwidth]{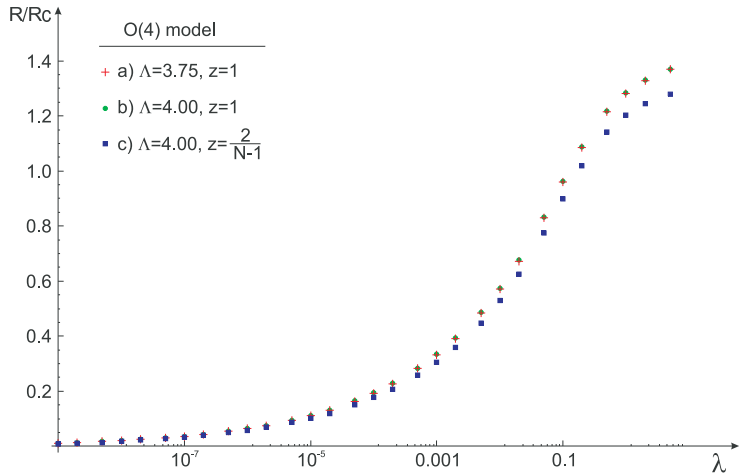}
\caption{Dependence of the mean radial field (in $R_c=m/\sqrt{\lambda}$ units) on the coupling constant $\lambda$ at zero temperature on a lattice $16^4$.} \label{Demchik:fig:1}
\end{figure}

In this paper we consider MC simulations at $N=4$, only. Other possible values of $N$ we tried lead to similar results. All the calculations are performed with the double precision. The system is thermalized by passing up to $10^{5}$ MC iterations for every run. For measuring, we use $10^{3}$ MC configurations separated by 10 bulk updates. To exclude lattice finite-size effects we perform simulations on lattices $16^4$ and $32^4$. The full correspondence of key results is obtained on these different lattices.

For MC simulations, we use the universal software environment {\bf QCDGPU} \cite{Demchik:2014dkq} that allows to perform simulations for a number of frequently studied models -- the $SU(N)$ gluodynamics and the $O(N)$ scalar field models. To produce pseudo-random numbers in {\bf QCDGPU} package we use our own library of pseudo-random number generators for MC simulations {\bf PRNGCL} \cite{PRNGCL:2014}.

All the simulations are performed on a heterogeneous distributed {\bf HGPU} cluster \cite{Demchik:hgpu}. It consists of different graphics processing units: AMD Radeon HD 7970, HD 6970, HD 5870, HD 5850, HD 4870, HD 4850, NVIDIA GeForce GTX 560 M, GTX 560 Ti.

The dependence of the averaged radial field $R/R_c$ on the coupling constant $\lambda$ at zero temperature on the symmetric lattice $16^4$ is shown in Fig. 1. Let us note that this quantity is not the scalar condensate value (the order parameter). In fact, it accounts for the scalar field coming from  the  classical condensate, quantum fluctuations and the part due to the logarithm of the Jacobian. The latter part is independent of the sign of the mass squared. In case of the classical condensate with no fluctuations, $R/R_c=1$ independently of $\lambda$. As we can see, $R/R_c$ is close to 1 for popular values of $\lambda\ge10^{-2}$, the quantum corrections to this quantity is no more than 40\%. However, $R/R_c$ goes to zero at extremely small couplings. At $\lambda < \la_0 \simeq 10^{-5}$ it is of one order less than the classical condensate value. At extremely small couplings the scalar field on a lattice exhibits tendency to be negligibly small with respect to the value of classical condensate. This result contradicts the idea of small quantum fluctuations over a strong homogeneous condensate used in our calculations. The same value of $\la_0$ is also detected on the lattice $32^4$ that demonstrates the stability of the results with respect to finite-volume effects. This dependence is not expected at the classical level where the condensate of $\phi$ has the value ${|m|}/{\sqrt{\la}}$.

We cannot interpret the decay of the homogeneous condensate on a lattice at $\lambda \to 0$ as some kind of symmetry restoration. In first, there is no evident saddle point to integrate the angular modes in spherical coordinates in case of restored symmetry, so, the restored phase is probably outside the scope of our effective model. In second, the classical potential becomes unstable at $\lambda=0$. In this regard, we can assume that the instability of the homogeneous condensate (or even instability of the $O(N)$ model itself) is reached at $\lambda\simeq 10^{-5}$.

The change of the SSB at extremely low couplings reminds in some aspects another phenomenon already found for continuous field theory by Linde \cite{Linde:1975sw} and Weinberg  \cite{Weinberg:1976pe} in the Higgs model and in the standards model, respectively. These authors observed that  the SSB does not happen for small values of the coupling $\la \leq \la_0$. Although the critical value  $\la_0$ depends on the mass value and gauge coupling $e^2$ value entering the Lagrangian, it is natural to consider small couplings to reach the Linde-Weinberg low bound. Physically, the Linde-Weinberg bound reflects the important property of the SSB -- the existence of the parameter ranges allowing the total effective potential to be dominated by the positive radiation quantum effects instead of the negative classical part. Another example of degeneration of the SSB is the scalar model at finite temperatures. Studying the phase transition in the $O(1)$ model on a lattice, we observed disappearance of the symmetry breaking at extremely small couplings \cite{Bordag:2012nh}. Although there is no direct correspondence between the scalar $O(N)$ model at zero temperature in the current paper and the mentioned examples of degeneration of the SSB, these analogies reflect general idea that the SSB in theories with scalar fields can change its behavior at some extremal values of coupling $\lambda$.

\section{Conclusions }
We have investigated the SSB phenomenon in the $O(4)$ model in the four-dimensional space-time on a lattice and determined its dependence on the coupling values. We have shown that symmetry breaking takes place for $\la \geq \la_0 \simeq 10^{-5}$. This is in the course of common belief that the SSB of continuous internal symmetries is a realistic mechanism for particle mass generation. It was experimentally grounded due to the recent discovery of the Higgs boson. At the same time, the standard model based on the $O(4)$ model is insufficient from both phenomenological and theoretical points of view and needs to be substituted by a model with extended scalar sector. For example, the Two-Higgs-Doublet standard model with $O(8)$ scalar sector is often discussed. In this regard, the investigation of the  dependence of SSB on $\la$ value can give a low bound on the scalar  particle masses for different $N$.

To our knowledge, the dependence on $\la$ value of the SSB and symmetry restoration at high temperature has not been investigated in detail  (except for paper \cite{Bordag:2012nh}) neither in $O(1)$ nor $O(N)$ models, yet. As a rule, several coupling values are selected on some reasons. For example, in \cite{Bimonte} the $O(4)$ model in $d = 3$ was studied with the normalized dimensionless coupling values chosen in the interval $5 \cdot 10^{-4} \leq \la \leq 10^{-2}$. In \cite{Agodi:1994qv} the coupling values $\la = 1, 1.5, 2$ were taken. In \cite{Jansen:1989gd} the limit $\la \to \infty$ has been investigated. In the literature we found a number of observations which are similar to what has been seen in the $O(1)$ model at finite temperature and in the $O(4)$ model above.  First of all, we note that in \cite{Agodi:1994qv} a close similarity of the $O(1)$ and $O(2)$ models was observed and qualitatively explained. It was shown that the Goldstone angular rotations can be integrated out and result in the common factor which is inessential for dynamics. So, symmetry behavior at finite temperature has to be similar in these models. We have also determined these properties of the Goldstone modes and described them in sect. 5. In  \cite{Bimonte}, \cite{Tetradis:1995br} hysteresis behavior was observed for small values of couplings. However, a detailed analysis of this effect was not given. As it was shown in \cite{Bordag:2012nh}, at finite temperature such a behavior is related with the change of the type of the phase transition, as a natural explanation of the hysteresis.
We would like to stress here that the dependence of the SSB on the coupling value is a nonperturbative effect which cannot be expected beforehand. This also concerns changing the kind of the temperature phase transition. In the literature on the topic, most results have been obtained firstly in various perturbation schemes including resummations series of Feynman diagrams. It was observed that the dependence on $\la$ disappears at all and perturbation theory in this parameter becomes not reliable at the critical temperature. So, other perturbation schemes (in particular, the expansion in $1/N$ for large $N$) have been used. In such investigations the dependence on the $\la$ value was not considered. The coupling was usually taken to be of order $\la \sim 0.01 - 0.1$, and a second order phase transition has been detected. Such a behavior is in agreement with our analysis for $O(1)$ model on a lattice for these coupling values.
The extremely small $\la$ values were not considered at all.

The main object introduced in the present paper is the effective Lagrangian for the radial field obtained by means of integration over angular continuous variables in the spherical coordinates in the internal space. It is convenient for further lattice investigations.
We applied the saddle-point method of integration. Some important features of the PF in the spherical coordinates
in the static limit were observed and taken into account. In particular, the completely different $\lambda$-dependence of
the PF is found for different signs of the mass term. In case of negative mass squared, we obtained singular behavior
in the limit $\lambda \to 0$, whereas for the `normal' mass term the PF is independent of
$\lambda$ at all. These properties are distinguishable for detecting both the SSB
and the $\lambda$-dependence of it.

To better understand the observed dependence of SSB on $\la$ value let us return back to the classical potential $V(\phi)$ given by \Ref{V}. If $m^2 > 0$, then symmetry breaking happens and the vacuum scalar condensate $\phi^2_c = {m^2}/{\la}$ is generated. In sect. 4, to investigate this phenomenon with accounting for quantum effects, we introduce the PF in the spherical coordinates \Ref{Demchik:Z}-\Ref{Demchik:Z1} and develop a modified saddle-point method of calculations. As important ingredient of it, a stationary point is determined with accounting for the Jacobian of the functional integral.  Actually,  this procedure defines the functional integration in the spherical coordinates. The value $R^2_c = \phi^2_c$ is used in actual calculations in sect. 6. So, we can discuss the dependence of the SSB on the $\la$ value in the frame of these definitions and procedures. For the `normal' mass term ($m^2 < 0$), the integral  \Ref{Demchik:Z1} converges and exists at $\la = 0$. But this is not the case for $m^2 > 0$ when the limit $\la \to 0$ does not commute with the integration. Moreover, the integral \Ref{Demchik:Z} does not exist in this limit that reflects the instability of the noninteracting tachyon modes at small momenta. For instance, this can be seen from the PF in the static limit (\ref{Demchik:z0}). The convergence of the functional integral requires $\la > 0$. At the same time, this is a necessary condition for condensation of the tachyon modes. In principle, the critical value of $\la_0$ can be estimated from the requirement that self-consistent quantum model exists. Physically, a classical scalar field is the condensate of tachyons generated due to self-interaction of them. If this interaction is very weak, quantum fluctuations can destroy the condensate and the lattice with properly chosen spacing could not detect it in the whole space. Formally, this does not exclude the formation of the condensate in small domains of the lattice, but this is a separate complicated problem beyond our analysis. Here we can assume that the homogeneous condensate inside the system becomes possible for $\la > \la_0$ and the dependence of the SSB on the $\la$ value is completely quantum nonperturbative effect.

It worth to remind that our results for pure scalar theory should not be directly identified with the known Linde-Weinberg low bound, which was observed in perturbation theory at one-loop order. In Refs. \cite{Linde:1975sw,Weinberg:1976pe} the models of interacting gauge and scalar fields were investigated and the lower limits on the Higgs particle mass was obtained. Disappearance of the SSB in those models is caused by different signs of the contributions of scalar and gauge fields to effective potential. So, strictly speaking, the Linde-Weinberg bound can be mentioned here as an example of critical values of coupling changing the picture of SSB.

The open-source software package {\bf QCDGPU} was used for MC simulations on GPU. We also propose the MC procedure for studying
the dependence of the SSB phenomenon on the coupling constant $\lambda$ values in
the $O(N)$ models. Within these facilities, we  detected that the scalar condensate has to disappear  at extremely small values of coupling,
$\lambda < 10^{-5}$ for the $O(4)$ model. This is in correspondence with the results obtained for the $O(1)$
model \cite{Bordag:2012nh}, where the critical value is also $\lambda_0 \simeq 10^{-5}$.  For $N > 1$ this behavior means that the SSB of continuous symmetries also depends on the $\la$  value.

In course of the results obtained, obvious problems for further studies are the $\la$-dependence of the SSB in the $O(N)$ models for $d = 1-3$ and the temperature phase transition. These investigations can be fulfilled on principles developed here and by using the worked out software package. It will be reported elsewhere.

\end{document}